\documentclass[twocolumn,showpacs,preprintnumbers,amsmath,amssymb]{revtex4}
\usepackage{amssymb}

\usepackage{graphicx}
\usepackage{bm}

\begin{document}

\title{Experimental Decoy Quantum Key Distribution
up to 130KM Fiber}

\author{Zhen-Qiang Yin, Zheng-Fu Han*, Wei Chen, Fang-Xing Xu, Guang-Can Guo}
\affiliation{Key Lab of Quantum Communication and Computation, CAS,
USTC, China }

\date{\today}

\begin{abstract}
Decoy State Quantum Key Distribution (QKD), being capable of beating
PNS attack and unconditionally secure, have become an attractive one
recently. But, in many QKD systems, disturbances of transmission
channel make quantum bit error rate (QBER) increase which limits
both security distance and key bit rate of real-life decoy state QKD
systems. We demonstrate the two-intensity decoy QKD with one-way
Faraday-Michelson phase modulation system, which is free of channel
disturbance and keeps interference fringe visibility (99\%) long
period, near 130KM single mode optical fiber in telecom (1550 nm)
wavelength. This is longest distance fiber decoy state QKD system
based on two intensity protocol.
\end{abstract}

\maketitle

\section{Introduction}

Quantum Key Distribution (QKD) \cite{BB84,ekert1991,Gisin}, as a
combination of quantum mechanics and cryptography, can help two
distant peers (Alice and Bob) share string of bits, called key. With
key and one time pad method, absolutely secure communication become
possible. However, most of QKD protocols, such as BB84, needs single
photon source, which is not practical for present technology.
Usually, real-file QKD set-ups
\cite{qkd1,qkd2,qkd3,qkd4,F-M,F-M2,single SPD} use attenuated laser
pulses (weak coherent states) instead. It means the density matrix
of states of photons emitted from Alice is:
$\rho=\sum_{n=0}^{\infty}\frac{\mu^n}{n!}|n\rangle\langle n|$.
Therefore, a few multi-photons pulses in the laser pulses emitted
from Alice opens the door of Photon-Number-Splitting attack (PNS
attack) \cite{PNS1,PNS2,PNS3}. Fortunately, decoy state QKD theory
\cite{decoy theory1,decoy theory2,decoy theory3,decoy theory4}, as a
good solution to beat PNS attack, has been proposed. The essential
idea of decoy state QKD is randomly changing the intensity (average
photon number) of the laser pulses from Alice, then Bob can get
different counting rates of laser pulses of different intensities.
From this, Alice and Bob can calculate the lower bound of counting
rate of single photon pulses ($S_1^L$) and upper bound of quantum
bit error rate (QBER) of bits generated by single photon pulses
($e_1^U$). At last, with error correction and privacy amplification,
unconditionally secure key could be get.

 Now, among protocols of decoy state QKD, two-intensity protocol
\cite{decoy theory4} and three-intensity protocol \cite{decoy
theory3} are ready for experiment. The former just uses two states:
coherent states with average photon number $\mu$, called signal
state, and $\nu$, called decoy state, satisfying $\mu>\nu$. $S_1^L$
and $e_1^U$ for two-intensity protocol are given by \cite{decoy
theory4}:
\begin{equation}
\begin{aligned}
S_1 &\ge S_1^L = \frac{\mu}{\mu\nu-\nu^2}(S_\nu^L e^{\nu}-S_\mu
e^\mu\frac{\nu^2}{\mu^2}-E_\mu S_\mu e^\mu
\frac{\mu^2-\nu^2}{\frac{1}{2}\mu^2})\\
e_1 &\le e_1^U = \frac{E_\mu S_\mu}{S_1^{L}\mu e^{-\mu}},
\end{aligned}
\end{equation}
where,
\begin{equation}
S_\nu^L = S_\nu(1-\frac{u_\alpha}{\sqrt{N_\nu S_\nu}}),
\end{equation}
Here $N_\nu$ is the number of pulses used as decoy states, $E_\mu$
is quantum bit error rate of $\mu$ laser pulses, $S_\mu$ is counting
rate of signal pulses, and $S_\nu$ is counting rate of decoy pules.
Therefore the lower bound of key generation rate ($R^L$) is:

\begin{equation}
R \ge R^L=q\{-S_\mu f(E_\mu)H_2(E_\mu)+S_1^L\mu
e^{-\mu}[1-H_2(e_1^U)]\}
\end{equation}
where, $f(E_\mu)$ represents bidirectional error correction
efficiency and q depends on implementation (1/2 for BB84 protocol).

 Recently, two-intensity protocol and three-intensity protocols have been
implemented in several experiments \cite{decoy experiment1,decoy
experiment2,decoy experiment3,decoy experiment4,decoy experiment5,
decoy experiment6}. In \cite{decoy experiment1, decoy experiment6}
two-intensity decoy QKD protocol was successfully performed, though
Plug\&Play system is not unconditionally secure. In \cite{decoy
experiment2}, a long distance (102KM) three-intensity decoy state
QKD experiment based on polarization modulation was demonstrated. In
\cite{decoy experiment3}, researchers finished a very long distance
(107KM) three-intensity decoy QKD, but their experiment used
ultra-low-noise, high efficiency transition-edge sensor
photo-detectors, which may be not very practical to most commercial
QKD systems.

 To prolong security distance of ordinary QKD or decoy state QKD,
depressing QBER is necessary. To keep stability of interference
fringe visibility is essential for depressing QBER, especially for
long distance case. In fact, polarization disturbances introduced by
quantum channel and optical devices is primary cause to decrease
interference fringe visibility and increase probability that a
photon hit the erroneous detector, which makes QBER rise. One way
Faraday-Michelson QKD system \cite{F-M,F-M2} can be free of the
disturbance of transmission fiber, to keep stability of interference
fringe visibility. Here, in our experiment, we have implemented
two-intensity decoy QKD experiment over 120KM single mode fibers,
just with one avalanche diode single photon detector (SPD).

 One SPD scheme \cite{single SPD} differs from traditional phase-modulation
type QKD system. In the latter, Bob randomly chooses between his
phase shifts $0$ or $\pi/2$, then Bob must use two SPDs to record
his photon counts. The two different phase shifts represent the two
conjugate bases of BB84 respectively, and one detector records bit
$0$, the other records bit $0$. However, in single SPD scheme, both
Alice and Bob choose between phase shifts $0$, $\pi/2$, $\pi$ and
$3\pi/2$. Alice and Bob just take phase shifts $0$ and $\pi/2$ as
bit $0$ and others as bit $1$. In fact, the only difference is that
in one SPD scheme Bob only detects phase difference of $0$ or $\pi$,
while in two SPDs scheme Bob detects phase difference of $0$ and
$\pi$. Though the counting rate of one SPD scheme is half of that of
two SPDs scheme, one SPD scheme may have security advantages over
two SPDs scheme. Vadim Makarov et al have proposed an attack to two
SPDs scheme, utilizing the detectors efficiency mismatch (see
\cite{attack1} for details). One SPD scheme is immune to this
attack. The use of optical circulators both in Alice and Bob makes
our system also immune to large pulse attack\cite{attack2,attack3}.

\section{Experiment Set-up}

 Our experiment set-up consists of control system, optical system,
synchronization light detector (SLD) and avalanche photon diode SPD
(just one SPD with dark counting rate $~5\times10^{-7}$). Based on
Faraday-Michelson phase modulation \cite{F-M}, the interference
visibility keeps high and consistent. Repetition frequency of our
system is 1MHz. The flow for an operation which means the process of
a laser pulse (decoy or signal) emitted form Alice and detected by
Bob is below:

 Alice randomly triggers the decoy or signal laser diode (DFB laser diodes) to emit decoy
laser pulse or signal laser pulse (quantum light for abbreviation)
and drives synchronization laser diode to emit synchronization laser
pulse at the same time. After emitted from Alice, quantum light
enters Alice's Faraday-Michelson interferometer, attenuated by
electrical variable optical attenuator (EVOA) to proper intensity
(average photon number per pulse: 0.6 for signal pulses, and 0.2 for
decoy signal pulses), enters 123KM single mode fiber (quantum
channel), phase-modulated by Bob's Faraday-Michelson interferometer
and is detected by Bob's SPD at last. Synchronization laser pulse
goes through another single mode fiber (synchronization channel)
which is almost as long as quantum channel. After emitted from
Alice, synchronization laser pulse enters synchronization fiber
immediately, in a while is detected by SLD, and then SLD gives a
signal to notify control board of Bob. Then Bob's control board
makes his phase modulator get ready for this operation, and after a
subtle delay, control board of Bob generates a trigger signal to
SPD, which detect the quantum light pulse and tell the result to
control board. After all operations finished, Alice announces decoy
and signal information and phase modulation information through
classical communication. According to this information, Bob
calculates $S_\mu$, $S_\nu$ and then $S_1^L$, $E_1^U$ through
equation (1) and (2). Now we can perform error correction and
privacy amplification to get unconditionally secure key. The
structure of our two-intensity decoy QKD system is demonstrated on
figure 1.

\begin{figure}[!h]
\center \resizebox{8.5cm}{!}{\includegraphics{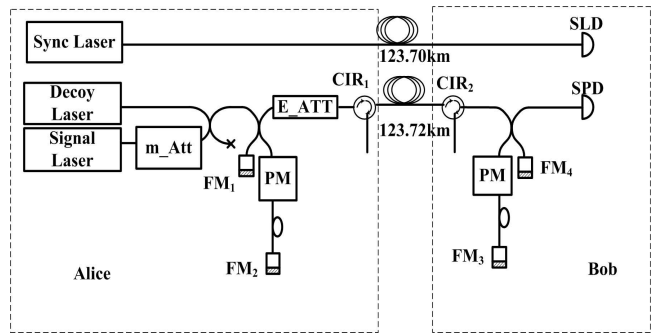}}
\caption{m\_Att: manual attenuator which modulates the intensity
ratio of signal laser pulse and decoy laser pulse; FM: Faraday
mirror; PM: phase modulator; E\_ATT: EVOA; SLD: synchronization
laser detector; SPD: single photon detector; $CIR_1$: Alice 's
optical circulator which only allow light leave Alice's security
zone , never allow light enter Alice's security zone; $CIR_2$: Bob
's optical circulator which only allow light enter Bob's security
zone, never allow light leave Bob's security zone; the two
circulators make our system immune to large pulse
attack;}\label{schematic}
\end{figure}

Intensity Modulation: How to realize laser pulse intensity
modulation is first step to perform decoy state QKD. Through making
simple modifications to the ordinary QKD system to realize intensity
modulation is very important to widen the use of decoy state QKD. In
our experiment, we use two laser diodes method to realize laser
pulse intensity modulation. We add a manual optical attenuator to
one of laser diode output. Then a fiber optical beam splitter is
used to couple the two laser output. We carefully adjust the manual
optical attenuator to make sure ratio of the two laser pulse
intensity is 1:3. Now, we can modulate intensity through selecting
different laser diode. With changing voltage on EVOA, we can also
modulate the intensity of laser pulse, but the repetition frequency
of EVOA is too low. Two laser diodes method is very convenient and
able to work with high repetition frequency.

Synchronization: Synchronization, especially to find the precision
delay between synchronization laser pulse and quantum light pulse is
very important to lower the QBER. The timing jitter of our SLD is
less than 500ps, while the gate-width of SPD is 2.5ns. So the QBER
caused by timing jitter is deeply depressed.

Phase Modulation: How to precisely determine the phase modulation
voltage is essential for lowering QBER. Because of environmental
disturbance, the phase modulation voltage may drift randomly. To
avoid the influence of this drift, we use active phase compensation
scheme. According to the half-wave voltage of Alice's phase
modulator, Alice can set her phase modulation voltage (for $0$,
$\pi/2$, $\pi$ and $3\pi/2$) definitely. Before transferring laser
pulses for generating key, Alice sets an arbitrary phase modulation
voltage, and then emits strong laser pulses to the quantum channel,
then Bob scans the whole possible phase modulation voltage and
watches the counting rates from SPD. According to results of this
scan, Bob can determine his phase modulation voltage (for $0$,
$\pi/2$, $\pi$ and $3\pi/2$). The time spent to determine phase
modulation working points relies on the drift speed of
interferometers. In common, the ratio between time spent to
determine phase modulation working points and the total working time
is below 5\%.

\section{Results\&Conclusion}

Experiment Results:  We set $f(E_\mu)=1.2$, average photon number
$\mu=0.6$ for signal laser pulses and $\nu=0.2$ for decoy laser
pulses. The ratio of decoy laser pulse number and signal laser pulse
number is 1:1, and 2G laser pulses was emitted in total. Table I is
the results for the experiment. With the experiment results,
equation (1), (2) and (3), we can get $S_1^L$, $e_1^U$, and $R_\mu$.
In table II, the length verse $S_1^L$, $e_1^U$, and $R_\mu$ are
given. In Figure 2, a graph on the length verse $R_\mu$ are given
too.

\begin{table}[!h]\center

\begin{tabular}{c c c c c}
\hline Length (KM)& $S_\mu$ & $E_\mu$ & $S_\nu$ & $E_\nu$ \\
\hline $123.6$  & $3.8\times10^{-5}$ & $0.0199$ & $1.36\times10^{-5}$  & $0.041$\\
\hline $108$    & $7.1\times10^{-5}$ & $0.016$ & $2.52\times10^{-5}$   & $0.027$\\
\hline $97$     & $1.24\times10^{-4}$ & $0.015$ &$4.3\times10^{-5}$ & $0.017$\\
\hline $83.7$   & $1.57\times10^{-4}$ & $0.0145$&$5.28\times10^{-5}$ & $0.019$\\
\hline $62.1$   & $2.88\times10^{-4}$ & $0.0108$&$1.08\times10^{-4}$ & $0.0225$\\
\hline $49.2$   & $8.6\times10^{-4}$ & $0.0103$&$2.9\times10^{-4}$ & $0.020$\\
 \hline

\end{tabular}

\caption {The length of fiber, counting rates of $\mu$ laser pulse
$S_\mu$, QBER of key generated from $\mu$ laser pulse $E_\mu$,
counting rates of $\nu$ laser pulse $S_\nu$ and QBER of key
generated from $\mu$ laser pulse $E_\nu$. This values are all
measured directly from experiment.}\label{yqerbound}
\end{table}

With the experiment results, equation (2) and (3), we can get
$S_1^L$, $e_1^U$, and $R_\mu$. In table II, the length verse
$S_1^L$, $e_1^U$, and $R_\mu$ are given. In Figure 2, a graph on the
length verse $R_\mu$ are given too.

\begin{table}[!h]\center

\begin{tabular}{c c c c}
\hline Length (KM)& $S_1^L$ & $e_1^U$ & $R_\mu$ \\
\hline $123.6$  & $3.78\times10^{-5}$ & $0.0607$ & $9.59\times10^{-7}$\\
\hline $108$    & $8.09\times10^{-5}$ & $0.0426$ & $4.89\times10^{-6}$\\
\hline $97$     & $1.41\times10^{-4}$ & $0.0399$ &$9.29\times10^{-6}$\\
\hline $83.7$   & $1.69\times10^{-5}$ & $0.0409$&$1.07\times10^{-5}$\\
\hline $62.1$   & $4.46\times10^{-4}$ & $0.0211$&$4.77\times10^{-5}$\\
\hline $49.2$   & $1.09\times10^{-3}$ & $0.0247$&$1.06\times10^{-4}$\\
 \hline

\end{tabular}

\caption {The length of fiber, counting rate of single photon laser
pulse $S_1^L$, QBER of key generated from single laser photon pulse
$e_1^U$, rate of generating secure key $R_\mu$. This values are all
calculated through equation (2)and (3) with parameters from Table
I.}\label{yqerbound}
\end{table}

\begin{figure}[!h]\center
\resizebox{7.5cm}{!}{\includegraphics{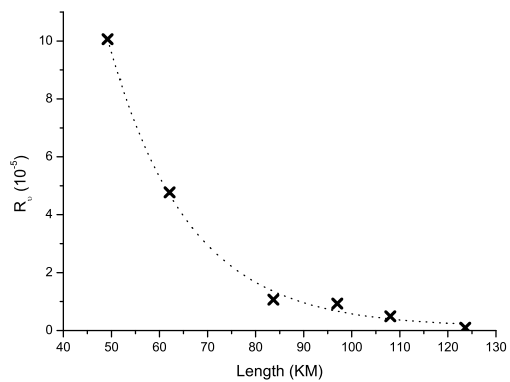}} \caption{Lower
bound of rate of secure key $R_\mu^L$ verse fiber length $L$. Each
point is get directly from experiment. }\label{schematic}
\end{figure}

 Form Figure 2, we find the limited fiber distance is about 130KM. We have successfully
realized up to 130KM decoy states QKD protocol just with simple
two-intensity protocol on one-way Faraday-Michelson phase modulation
system. And really unconditionally secure key can be distributed
through such a long distance fiber.

In conclusion, we have implemented two-intensity decoy QKD protocol
on the one-way Faraday-Michelson phase modulation QKD system with a
popular avalanche photon diode detector. Unlike many other QKD
systems which is suffered of disturbances of transmission channel,
one way Faraday-Michelson QKD system, which is free of polarization
disturbances caused by quantum channel and optical devices in the
system, can really keep steady and high interference fringe
visibility, and leads to low QBER. With low and steady QBER, both
security distance and key bit rate of decoy state QKD are improved.
It's noticeable that one way Faraday-Michelson QKD system free of
channel disturbances can be used directly in commercial condition
not only in lab. Our system can provide unconditionally secure key
distribution service up to 130KM optical fiber on telecom wavelength
(1550nm). So far, this distance is longest in real-life
two-intensity decoy state QKD systems.

 The authors thank Dr.Chi-Hang Fred Fung and Prof.Hoi-Kwong Lo for
reading the manuscript and helpful advice. This work was supported
by National Fundamental Research Program of China (2006CB921900),
National Natural Science Foundation of China (60537020,60621064) and
the Innovation Funds of Chinese Academy of Sciences. To whom
correspondence should be addressed, Email: zfhan@ustc.edu.cn.

\end{document}